\documentclass{PoS}

\title{Top Quark Production at Hadron Colliders: an Overview}

\ShortTitle{Top Quark Production at Hadron Colliders}

\author{R.~Bonciani\\
        Laboratoire de Physique Subatomique et de Cosmologie,
Universit\'e Joseph Fourier/CNRS-IN2P3/INPG,
F-38026 Grenoble, France\\
        E-mail: \email{bonciani@lpsc.in2p3.fr}}

\author{\speaker{A.~Ferroglia}
\\
        Institut f\"ur Physik (THEP), Johannes Gutenberg-Universit\"at,
D-55099 Mainz, Germany \\
        E-mail: \email{ferroglia@thep.physik.uni-mainz.de}}

\abstract{The study of the properties of the top quark is one of the main goals
of the Large Hadron Collider  (LHC) physics program. The experimental precision
expected at the LHC requires the calculation of several top-quark related
observables  beyond leading order  in the strong coupling constant. In this
work we briefly review the status of the theoretical predictions for the
top-quark production processes at hadron colliders. Special attention is
devoted to recent progress in the calculation of next-to-next-to-leading-order
corrections to the top-quark pair production cross section.}

\FullConference{European Physical Society Europhysics Conference on High Energy Physics,
EPS-HEP 2009,\\ 
July 16 - 22 2009\\
Krakow, Poland}

\newcommand{\be}{\begin{equation}}
\newcommand{\ee}{\end{equation}}
\newcommand{\bea}{\begin{eqnarray}}
\newcommand{\eea}{\end{eqnarray}}
\newcommand{\bd}{\begin{displaymath}}
\newcommand{\ed}{\end{displaymath}}

\newcommand{\bc}{\begin{center}}
\newcommand{\ec}{\end{center}}

\begin{document}

\section{Introduction}

With a mass $m_t = 173.1 \pm 1.3$ GeV, the top quark is the heaviest 
elementary\footnote{at least according to the Standard Model (SM) and its most
popular extensions, such as SUSY models.} particle  produced at colliders. Due
to their very large mass, top quarks have the unique property of decaying
through weak interactions before they can form bound states.
For the same reason, top quarks should couple strongly with the particles
responsible for the electroweak symmetry breaking and they
are expected to play a key role in the search for the Higgs boson and in
investigating the origin of particle masses. 
To date the properties and quantum numbers of the top quark could be
experimentally studied only at the Tevatron, where a few thousand  top quarks
were produced since the particle discovery in 1995. The situation will change
drastically when the LHC will start operations: the new proton-proton collider
is expected  to produce millions of top quarks per year already in the first 
low luminosity phase (${\mathcal L} \sim 10 \, \mbox{fb}^{-1}/\mbox{year}$).
The large numbers of events involving top quarks at the LHC will allow
precise measurements of 
their properties and production cross sections.
Precise measurements demand  for  equally precise theoretical predictions of
the measured  observables in the Standard Model (SM) and its extensions.
Precise theoretical predictions are obtained by pushing the calculation of the
physical observables to the next-to-leading-order (NLO) or,  if necessary, even
to the next-to-next-to-leading-order (NNLO) in perturbation  theory, as well as
by applying resummation techniques in specific phase-space  regions.
Top quarks appear in processes studied at colliders as virtual particles.
However, due to the small ratio between the top-quark width and mass,  it is
possible to factor the cross section of processes involving top quarks into the
product of the production cross section for on-shell top quarks and the
top-quark decay width. The purpose of this short write-up is to review the
status of the theoretical predictions for the top-quark production cross
sections relevant at the LHC. We will focus primarily on inclusive cross
sections. For a comprehensive (and excellent) review of  many other aspects of 
top-quark phenomenology, see \cite{Bernreuther:2008ju}.

\section{Top-Quark Pair Production \label{toppair}}

At hadron colliders top quarks are primarily produced in pairs with their
antiparticle. In about $5 \cdot 10^{-25}$ seconds the  top (antitop) quark
decays in a $W$-boson and a $b$-quark ($\bar{b}$-quark). The $W$-boson can
either decay leptonically or give origin to a pair of light-quark jets. 

The production of an on-shell top-antitop pair is dominated by  strong 
interactions.  The inclusive top-quark pair production  cross section can be 
written as  
\be
\sigma^{t \bar{t}}_{h_1,h_2}(s_{\mbox{{\tiny had}}},m_t^2) = \sum_{ij} \int_{4
m_t^2}^{s_{\mbox{{\tiny had}}}} d \hat{s}\,  \underbrace{L_{ij}\left(\hat{s},
s_{\mbox{{\tiny had}}}, \mu_f^2 \right)}_{\mbox{{\footnotesize partonic
luminosity}}}\, 
\overbrace{\hat{\sigma}_{ij}(\hat{s},m_t^2,\mu_f^2,\mu_r^2)}^{\mbox{{\footnotesize
partonic cross section}}}  \, , 
\label{CS}
\ee  
where the hard scattering of the partons $i$ and $j$  ($i,j \in \{q , \bar{q},
g\}$) at a partonic center of mass (c.~m.) energy $\hat{s}$ is described by the
partonic cross section, which can be  calculated in perturbative QCD. The
process independent partonic luminosity describes the probability  of finding,
in the hadrons $h_1$ and $h_2$ ($h_1, h_2 = p, \bar{p}$ at the Tevatron, $h_1,
h_2 = p, p$ at the LHC), an initial state involving partons $i$ and $j$ with
the given partonic energy  $\hat{s}$. The integration extends up to the
hadronic c. m. energy $s_{\mbox{{\tiny had}}}$ ($s_{\mbox{{\tiny had}}} = 1.96$
TeV at the Tevatron, $s_{\mbox{{\tiny had}}} = 14$ TeV at the LHC).  The
renormalization and factorization scales, $\mu_f$ and $\mu_r$,  are usually set
to be equal.

There are two production channels contributing to the partonic cross section at
the tree level: the quark-antiquark channel $q \bar{q} \to t \bar{t}$ and the 
gluon fusion channel $gg  \to t \bar{t}$.
Because of the interplay between parton luminosity and  partonic cross
sections, the quark-antiquark channel dominates at the Tevatron, where  it
gives origin to $\sim 85 \%$ of the inclusive cross section.  On the contrary,
at the LHC the inclusive cross section is largely dominated by  gluon fusion
events, which contribute $\sim 90 \%$ of the total cross section.
NLO QCD corrections to both channels have been known for two decades
\cite{Nason:1987xz} and were recently obtained in analytic form 
\cite{Czakon:2008ii}; these corrections are very large (see Fig.~\ref{tabX}).
Mixed QCD-EW corrections are also available \cite{QCDEW}, but they have a 
negligible impact on the inclusive cross section.
NLO QCD effects involve logarithmic terms which become numerically sizable near
the production threshold. Up to date theoretical predictions for the top-quark
pair-production cross section include effects originating from the resummation
of these logarithms (see Section~\ref{resummation}), which are particularly
relevant at the Tevatron, where the bulk of the inclusive cross section  arises
from events with a partonic c. m. energy close to the production threshold 
$\sqrt{\hat{s}} \sim 2 m_t$.
The inclusive cross section is sensitive to the top-quark mass ($\Delta
\sigma^{t \bar{t}} / \sigma^{t \bar{t}} \sim - 5 \Delta m_t/ m_t$). Current
theoretical predictions (including NLO corrections and
next-to-leading-logarithm resummation) indicate a  cross section  $\sigma^{t
\bar{t}}  \sim 7.6$ pb with a relative uncertainty  of $\sim 11 \%$ at the
Tevatron and $\sigma^{t \bar{t}}  \sim 908$ pb with a  relative uncertainty of
$\sim 13 \%$ at the LHC, for $m_t \sim 171$ GeV  \cite{Cacciari:2008zb}.
These results are obtained using the CTEQ6.5 set of PDFs \cite{CTEQ6.5}.  Using
the set MRST2006nnlo \cite{MRST}, a sizable difference is observed. 
Calculations including approximated NNLO corrections 
\cite{Moch:2008qy,Kidonakis:2008mu} indicate smaller uncertainties.
At the Tevatron, as at the LHC, the theoretical uncertainty is dominated by
unknown higher order corrections, whose magnitude is estimated through the
residual scale dependence of the cross section.  The uncertainty affecting the
parton luminosity is the other source  of the theoretical uncertainty.
The theoretical uncertainty should be
compared with the expected accuracy of the experimental measurements at the
LHC, where the inclusive top-quark pair production cross section is likely to
be measured with a relative error of $5-10 \%$ \cite{Bernreuther:2008ju}. Such
precise measurements demand for equally precise  cross-section  predictions,
which can be achieved only by including the effects of the NNLO QCD
corrections (see Section~\ref{NNLO}).

The sensitivity of the inclusive cross section to the top-quark mass allows one
to obtain from the cross-section
measurement a numerical value for a  well-defined short distance mass, such as
the running $\overline{\mbox{{\footnotesize MS}}}$ mass. (This is not possible with the current mass measurements based on
kinematic reconstruction of the events.)  In \cite{Langenfeld:2009tc}, by
employing the approximated NNLO corrections of
\cite{Moch:2008qy,Langenfeld:2009tc}  and the cross-section measurement 
\mbox{$\sigma^{t \bar{t}}_{\mbox{\tiny EXP}} = 8.18^{+0.98}_{-0.87}$ pb} 
\cite{Abazov:2009ae}, the $\overline{\mbox{{\footnotesize MS}}}$ mass  of the
top quark was  calculated  to be equal to  $\overline{m}_t(m_t) =
160.0^{+3.3}_{-3.2}$ GeV.

\begin{figure}[th]
\begin{center}
\begin{tabular}{|c||c|c|c|c|}
\hline
    LHC ($s_{\mbox{{\tiny had}}} = 14$ TeV) & $\sigma^{t \bar{t}}$ & $\sigma^t$ $t-$channel &
$\sigma^t$ $s-$channel & $\sigma^t$ associated $t W$ \\
\hline
\hline
NLO QCD & $\sim +50 \%^\S$  & $\sim +5\%^\flat$& $\sim +44 \%^\bot$ &
$\sim +10 \%^\top$ \\ 
EW & $\sim -0.5\%^\ddag$&$< 1\%^\natural$ & & \\ 
MSSM &up to $\pm 5 \%^\P$ &$< 1\%^\natural$ & &  \\ 
\hline
\end{tabular}
\vspace*{1mm}\\
\begin{tabular}{|c||c|c|c|}
\hline
Tevatron ($s_{\mbox{{\tiny had}}} = 1.96$ TeV) & $\sigma^{t \bar{t}}$ & $\sigma^t$ $t-$channel &
$\sigma^t$ $s-$channel  \\
\hline
\hline
NLO QCD & $\sim +25 \%^\S$  & $\sim +9\%^\flat$&$\sim +47 \%^\bot$  \\ 
EW & $\sim -1\%^\ddag$ & $< 1\%^\natural$ &  \\ 
MSSM & up to $\pm 5 \%^\P$&$< 1\%^\natural$ &  \\ 
\hline
\end{tabular}
\end{center}
\vspace*{-3mm}
\caption{Approximate size of the NLO QCD and MSSM corrections to
pair and single-top production cross sections with respect to the Born cross 
sections. The size of the electroweak (EW)  corrections is expressed in  $\%$
of the NLO QCD cross sections.  References: $^\S$ \cite{Nason:1987xz}; $^\P$
\cite{Berge:2007dz}; $\ddag$ \cite{QCDEW}; $\flat$ \cite{Bordes:1994ki};  
$\natural$\cite{Beccaria:2006ir}; $\bot$ \cite{SW}  ; $\top$ \cite{GZ}.
\label{tabX}}
\vspace*{-2mm}
\end{figure}

\subsection{Soft-Gluon Resummation \label{resummation}}

The  calculation of any physical observable in perturbative QCD is naively 
organized in  terms of a series of increasing powers of the coupling constant
$\alpha_S$. Formally, the  $n$th-order corrections are suppressed with respect
to the corrections of order $n-1$ by a power of $\alpha_S$ and, therefore, they
are sub-leading. However,
in particular regions of the phase space this classification can fail, and 
corrections of higher order in  $\alpha_S$ can be of the same numerical size as
the leading ones.

The QCD corrections to processes involving at least two large energy scales are
characterized by a logarithmic behavior in the vicinity of the boundary of the
phase space. This is precisely  the case of $t \bar{t}$ production, where
$\hat{s}$ and  $m_t^2$ are such that $\hat{s},m_t^2 \gg \Lambda_{QCD}^2$.
Let $\rho$ be the inelasticity variable of a certain process and let us
consider the quasi-elastic limit in  which $\rho \to 1$. The physical
observable, say  $\sigma$, will exhibit the following logarithmic behavior:
$\sigma \sim \sum_{n,m} C_{n,m} \alpha_S^n \ln^m{(1-\rho)}$ with $m \leq 2n$.
These logarithms come from the integration over the phase space of the plus
distributions originating from the cancellation  of the IR singularities in the
sum of virtual and real corrections in inclusive observables. They are,
therefore, a reminder of the IR divergences. Although  the physical observable
is formally finite, in the limit $\rho \to 1$ the virtual and real radiation
are unbalanced, due to the phase space restrictions, and the terms $\alpha_S^n
\ln^m{(1-\rho)}$ can become large, even in the perturbative regime, where
$\alpha_S \ll 1$.  The logarithmic terms spoil the convergence of the 
perturbative series and they must be resummed  to all orders. The resummation
can be carried out by  using the general factorization properties of QCD and
the behavior of the process-dependent phase space in the vicinity of the
elastic region, in order to re-express the multi-gluon amplitude in terms of
single soft gluon emissions. This allows an exponentiation of the logarithmic
terms.
The exponentiation is usually not possible in the $\rho$ space, but it takes
place in a conjugate space, for instance the Mellin space of $N$ moments, where
$N$ is the variable conjugate to $\rho$. It can be proved that, in the
conjugate space, the partonic cross section can be written using the
generalized resummation formula \cite{Sterman:1986aj}:
\be
\sigma^{res}_{N} \sim
\exp \bigl\{ \ln{N} \, g_{1}(\alpha_{S} \ln{N}) 
+ g_{2} \bigl(\alpha_{S} \ln{N}, Q^{2}/\mu^{2} \bigr) 
+ \alpha_{S} \, g_{3} \bigl( \alpha_{S} \ln{N}, Q^{2}/\mu^{2} \bigr) 
+... \bigr\} \, .
\label{e62}
\ee
In Eq.~(\ref{e62}) the function $g_{1}$ resums  the {\em leading  logarithms}
(LL). The function $g_{2}$, which is formally suppressed by a  power of
$\ln{N}$ with respect to $g_{1}$, resums the {\em next-to-leading  logarithms}
(NLL), $g_{3}$ the NNLL ones, and so on.
The resummation procedure outlined above applies to the production of a 
heavy-quark pair near the production threshold ($\rho = 4m_t^2/\hat{s}$) 
\cite{LL,NLL}. 
In this case, the soft radiation is emitted by both the initial and the final
state, and color correlations make in such a way that a trivial $c$-number
exponentiation of the form shown in Eq.~(\ref{e62}) is, in general, no longer
correct. However, in the threshold limit, it can be shown that, at the NLL
level, the final $t \bar{t}$ state acts, with respect to the soft radiation, as
a single colored particle, either in the singlet or in the octet state.
Consequently, one can recover a $c$-number exponentiation as in 
Eq.~(\ref{e62}) independently for the singlet and octet states.
Recent studies considered the extension of the resummation for the $t \bar{t}$
production cross section at the NNLL level \cite{Moch:2008qy,NNLL}.
The resummed cross section has a smaller dependence on the 
renormalization/factori\-zation scale in the vicinity of the elastic region. In
the case of $t \bar{t}$ production at LHC, the central value of the  NLO+NLL
cross section  increases with respect to the pure NLO result  by about +4\%, 
while the dependence on the factorization/renormalization scale variation is
reduced by a couple of percents.
At the Tevatron, while the  central
value of the cross section is increased by the same percentage as at the LHC,
the dependence on  the factorization/renormalization scale  of the NLO+NLL
cross section is roughly half of the one found in a pure NLO calculation.

\subsection{NNLO Corrections \label{NNLO}}

The NNLO QCD corrections to the partonic pair-production cross section can be
sorted in different sets as follows: {\em Virtual corrections}, which include
{\em i-a)} two-loop corrections to the processes  $q \bar{q} \to t \bar{t}$
and  $g g \to t \bar{t}$, to be interfered with the tree-level amplitude, and
{\em i-b)} one-loop diagrams for the quark-antiquark and gluon fusion channels,
to be interfered among themselves; {\em Real corrections}, including {\em
ii-a)} one-loop diagrams with one extra parton in the final state, to be
interfered with the corresponding tree-level amplitude, and  {\em ii-b)}
tree-level diagrams with a top-quark pair plus two extra partons in the final
state.
The Feynman diagrams belonging to the sets {\em ii-a)} and {\em ii-b)} were
evaluated in the context of the calculation of the NLO corrections
for the production of a top-quark pair plus one jet \cite{ttj}.
The implementation of these results in the inclusive calculation of the 
$t \bar{t}$ production cross section will require an extension of the 
available subtraction methods at the NNLO.
The calculation of the interference of one-loop graphs (set {\em i-b)}) was 
completed last year \cite{oneXone}. 
The two-loop corrections belonging to the set {\em i-a)} are known in the 
limit in which the top-quark mass is considered much smaller that the partonic
c. m. energy \cite{smallmass}. The latter results, which were obtained by
employing the factorization theorem proposed in \cite{fact}, are not sufficient
for phenomenological applications, since events with a partonic c. m. energy
close to the production threshold make up a sizable fraction of the inclusive
pair-production cross section.  In the quark-antiquark channel, the complete
set of two-loop corrections was calculated numerically in \cite{Czakon:2008zk}
for arbitrary values of the Mandelstam invariants and of the top-quark mass.
The technique employed was based on the reduction of the squared amplitude to
Master Integrals (MIs) by means of the Laporta algorithm, followed by a
numerical solution of the differential equations satisfied by the MIs (see
references in \cite{Bonciani:2008az, Bonciani:2009nb}). All the two-loop
diagrams involving a closed massive or massless quark loop in the $q \bar{q}$
channel were evaluated analytically in \cite{Bonciani:2008az}. The calculation
was carried out by employing the Laporta algorithm and by solving analytically
the differential equations  satisfied by the MIs. With the same technique it
was  possible to calculate the leading color coefficient in the interference of
the two-loop corrections to  $q \bar{q} \to t \bar{t}$ with the tree-level
amplitude \cite{Bonciani:2009nb}. 

Exact results for the two-loop corrections to the gluon fusion channel are not
yet available. However, the analytic expression of the  infrared (IR) poles in
the interference of the two-loop gluon-fusion diagrams  with the tree-level
amplitude is known \cite{Ferroglia:2009ii}. These poles were calculated by
employing the expression for the IR poles for a generic two-loop amplitude in
massive QCD derived in \cite{Becher:2009kw}. For what concerns the finite
parts, it seems  possible to calculate analytically  part of the two-loop
diagrams in the gluon fusion channel by employing the same techniques applied
to the calculation of the two-loop corrections to the quark-antiquark channel.
However, it is known that several two-loop diagrams (for example some box
diagrams involving a closed heavy quark loop) cannot be expressed in terms of 
generalization of harmonic polylogarithms. For this kind of diagrams, a
numerical approach could be unavoidable.

\section{Single Top Quark Production\label{singletop}}
 
\begin{figure}[t] 
\begin{center}
\begin{tabular}{|c|c|c|c|}
\hline
cross section &  $t$-channel ($pb$) & $s$-channel ($pb$) & 
$tW$ mode ($pb$) \\ 
\hline
\hline
$\sigma^t_{\mbox{{\tiny Tevatron}}}$ &$1.15 \pm 0.07$ & $0.54 \pm 0.04$ & $0.14 \pm 0.03$\\
\hline
$\sigma^t_{\mbox{{\tiny LHC}}}$ & $150 \pm 6$ & $7.8 \pm 0.7$ & $44 \pm 5$\\
$\sigma^{\bar{t}}_{\mbox{{\tiny LHC}}}$ & $92 \pm 4$ & $4.3 \pm 0.3$&$44 \pm 5$\\
\hline
\end{tabular}
\end{center}
\caption{Predictions for the single top and antitop production cross sections at 
the Tevatron and at the LHC for $m_t = 171.4 \pm 2.1$ GeV \cite{singletop}.}
\label{sintop}
\vspace*{-2mm}
\end{figure}

Single top quarks can be produced by flavor-changing weak interactions.  Like
the pair-production cross section, the single-top production cross section can
be factored in the convolution of a hard partonic process and a parton
luminosity. In the SM there are three main production channels relevant at the
Tevatron and at the LHC: {\em i)} the process $q (\bar{q}) b \to q' (\bar{q}')
t$ in which a $W$ boson is exchanged in the  $t$-channel, {\em ii)} the mode $q
\bar{q} \to \bar{b} t$, in which a  $W$-boson is exchanged in the $s$-channel,
and {\em iii)} the process $bg \to W^- t$, referred to as $tW$ associated
production channel. The tree-level SM cross sections for the three channels are
all proportional to the CKM matrix element $|V_{tb}|^2$. The single-top
production cross section are large enough to allow for millions of single top
events at the LHC. However, precise measurements are made difficult by the
large backgrounds. Tevatron experiments recently reported evidence for single
top production \cite{EXPsintop}. At the LHC it should be possible to measure
separately the cross section in the three different channels. 
The predicted value for the single-top production cross section in the  three
channels at the Tevatron and at the LHC are shown in Fig.~\ref{sintop}
\cite{singletop,Bernreuther:2008ju}. These predictions include NLO QCD
corrections \cite{Bordes:1994ki, SW, GZ}.
At the LHC, the $t$-channel cross section is expected to be measured with a
relative error of $10 \%$, while the other two channels will be measured with
larger uncertainties. It is therefore possible to conclude that the
theoretical  uncertainty on the single-top production cross sections is
currently under control.  \\

\noindent {\bf Acknowledgments}:
\noindent We thank T.~Gehrmann and C.~Studerus for 
several useful discussions and for a careful reading of the manuscript.
 

\end{document}